\documentclass[prd,12pt,english]{revtex4}
 \usepackage{babel}
\usepackage{amsmath} 
\usepackage{amssymb} 
\usepackage{graphicx} 
\usepackage{graphics} 
\usepackage{epsfig} 
\usepackage{slashed}
\usepackage{multirow}

\newcommand\nn{\nonumber} \newcommand\ba{\begin{eqnarray}}
\newcommand\ea{\end{eqnarray}} \newcommand\alb{\begin{align}}
\newcommand\ale{\end{align}} \newcommand\be{\begin{equation}}
\newcommand\ee{\end{equation}}

\newcommand{\thg}{\theta_{\gamma}}

\def\epem 	{\ensuremath{e^+ e^-}} 
\def\pp {\ensuremath{\bar p p }} 

\begin{document} 
\title{New fit of time-like  proton electromagnetic form
factors from  $e^+e^-$ colliders}

\author{Egle Tomasi-Gustafsson} \email{egle.tomasi@cea.fr}
\affiliation{\it CEA, IRFU, DPhN, Universit\'e Paris-Saclay, 91191
Gif-sur-Yvette Cedex, France}

\author{Andrea Bianconi} \email{andrea.bianconi@unibs.it}
\affiliation{\it Dipartimento di Ingegneria
dell\!~$^\prime$Informazione, Universit\`a degli Studi di Brescia and
Istituto Nazionale di Fisica Nucleare, Gruppo Collegato di Brescia,
I-25133, Brescia, Italy}

\author{Simone Pacetti } \email{simone.pacetti@unipg.it}
\affiliation{\it Dipartimento di Fisica e Geologia, and INFN Sezione di
Perugia, 06123 Perugia, Italy }

\begin{abstract} The data on the proton form factors in the time-like region
from the BaBar, BESIII and CMD-3 Collaborations are examined to have
coherent pieces of information on the proton structure. Oscillations in
the annihilation cross section, previously observed, are determined with
better precision. The moduli of the individual form factors, determined
for the first time,  their ratio and the angular asymmetry of the
annihilation reaction $\epem\to\pp$ are discussed. Fiits of the
available data on the cross section, the effective form factor, and the
form factor ratio, allow to propose a description of the electric and
magnetic time-like form factors from the  threshold up to the highest momenta.

\end{abstract}

\maketitle 

\section{Introduction}

The understanding  of the proton electromagnetic form factors (FFs),
called electric $G_E(q^2)$ and magnetic $G_M(q^2)$ Sachs FFs is the aim
of theoretical and experimental studies since decades, in the frame of a
unified view of the scattering and annihilation regions. Much progress
has been done recently, due, on one side, to new experiments that
collected information with better precision and/or in a wider
kinematical range and, on the other side, to theoretical efforts that
extend models and parametrizations built in the space-like (SL) region
to the time-like (TL) region (for a review, see Ref. \cite{Pacetti:2015iqa}).
We discuss here the data on  the $\epem\to\pp$
cross section, $\sigma_{\epem\to\pp}$, from the BaBar, BESIII and CMD-3 Collaborations, obtained
either by direct measurements of the annihilation process, or by means
of the so-called initial state radiation (ISR) technique, $i.e.$, by
exploiting the three-body process $\epem\to\pp\gamma$, where the photon
is radiated by one of the initial leptons.

The emission of a real hard photon, leaving the radiating lepton in a
``quasi-real'' state, allows extracting the cross section for the
process $\epem\to\pp$ from the  differential cross section of the
three-body process $\epem\to\pp\gamma$.  In such a kinematic domain,
$\sigma_{\epem\to\pp}$ factorizes out in the expression of the ISR
differential cross section. In collinear kinematics,  the ISR cross
section manifests a logarithmic enhancement as a consequence of the
small mass of the virtual electron that is almost on mass
shell~\cite{Baier:1973ms}. At fixed energy colliders the ISR technique
allows to extract values of the $\sigma_{\epem\to\pp}$ cross section at
different transferred momenta, $i.e.,$ different values of $q^2$ (being
$q$ the four-momentum of the virtual photon in the annihilation reaction
$\epem\to\pp$) by tuning the kinematics of the real photon. The cost is
a reduction of a factor of $\alpha
=e^2/(4\pi)\simeq 1/137 $ (the electromagnetic fine constant)
of the number of events, that, however, can be compensated by the high
luminosity recently achieved at the experimental facilities.

By means of the ISR technique and detecting the radiated hard photon,
the BaBar Collaboration obtained data on the $\epem\to\pp$ cross
section with an error lower than 10\% in a wide energy region, from the
production threshold $\sqrt{s}=2m_p$ up to
$\sqrt{s}~\simeq~6$~GeV~\cite{Lees:2013xe}, where $s=q^2$ is the total energy squared 
in the center of mass (CM) frame of the $\bar p 
p$-system and $m_p$ is
the proton mass. Recently, using the same technique but with undetected
initial photon, the BESIII Collaboration extracted 30 values of
$\sigma_{\epem\to\pp}$ in the range $(2\le\sqrt{s}\le
3.8)$~GeV~\cite{Ablikim:2019njl}.  The ISR photon is undetected, $i.e.$,
it is mostly emitted at small polar angles in a kinematical region
uncovered by BESIII acceptance. This method was also used by the BaBar
Collaboration, where the hard condition of the photon was insured by
high energy of the colliding beams \protect\cite{Lees:2013uta}.

The individual determination of the moduli of the FFs in the TL region
was done by the BESIII Collaboration, using the energy scan
method~\cite{Ablikim:2019eau}, with a  precision comparable to that of
the data obtained in the SL scattering region. The data in the
SL region were mostly collected by the JLab GEp Collaboration and published in a series 
of papers, summarized in Refs. \cite{Puckett:2017flj,Puckett:2011xg}.  
The BESIII Collaboration has made the individual measurement of $|G_E|$ and
$|G_M|$, separately, in the TL region for the first time ever.

In fact, before such a pioneering measurement, the few information on
the FF moduli in the TL region concerned a composed observable, namely
their ratio $R=|G_E|/|G_M|$, extracted from  angular distribution measurements.
Due to luminosity limitations, only an 'effective form factor' could be
extracted from the total cross section.

Let us stress that only the moduli of the FFs, which, in principle, have
a non-vanishing imaginary part in the TL region, can be extracted from a precise large-statistics measurement of the angular distribution of the final-state nucleons in the $e^+e^-$-CM frame.  
The underlying assumption is that the
reaction occurs through the one-photon exchange mechanism~\cite{Zichichi:1962ni}. No
measurement of the relative phase between $G_E$ and $G_M$, accessible through polarization
observables \cite{Dubnickova:1992ii}, is available yet for protons and
neutrons .

Focussed on the threshold region, the CMD-3 Collaboration \cite{CMD-3:2018kql} measured the cross section for the
reactions $  \epem \to \bar pp\gamma $ 	and $ \epem\to \bar
nn\gamma $. The scan of the nucleon-antinucleon threshold energy region is done by 
measuring  the beam energy at 0.1 MeV precision by back-scattering laser
light system. The energy spread due to radiation and energy resolution
is small enough to differentiate the proton and neutron thresholds.

The aim of the present work is to scrutinize the recent data  on proton
FFs in TL region, through the reaction $\epem\to \pp(\gamma)$. 
Two characteristics, earlier predicted or highlighted, can be confirmed
or infirmed by the new data: the finding of regular oscillations of the
cross section \cite{Bianconi:2015owa} and the steeper $q^2$-dependence of the
electric FF ($G_E$) compared to the magnetic FF ($G_M$), as found in the
SL region \cite{Jones:1999rz,Puckett:2017flj}. The suggestion of a similar
$q^2$-dependence in space and TL regions is based on analytical
properties of the amplitudes  \cite{TomasiGustafsson:2001za} and
illustrated in frame of a generalized definition of FFs
\cite{Kuraev:2011vq}.

Not all models developed in the SL region have the correct analytical
properties to be extended in the TL region,  where FFs are of complex
nature \cite{TomasiGustafsson:2005kc}. Models based on dispersion
relations \cite{Belushkin:2006qa}  or vector dominance
\cite{Bijker:2004yu,Lomon:2012pn} have attempted a global description in
SL and TL regions, for a review, see Refs. \cite{Pacetti:2015iqa,Denig:2012by}.
 In this paper we consider  the new data and we propose a global fit
from threshold up to the maximum available transferred momentum. The
individual TL FFs are reproduced from a fit on the ratio $R$ and of the
effective FF, allowing to extrapolate their behavior at threshold,
where $R$ is constrained to unity.

\section{ The $\epem\to\pp (\gamma)$ cross section }
As already pointed out, at fixed-energy $\epem$-colliders,
the $ \epem\to \pp $ cross section, 
can be extracted from the data on the differential
cross section of the ISR process $\epem \to \pp  \gamma$, 
where the photon is radiated by one of the initial electrons, 
over a range of $\pp $-energies going from the threshold,
$\sqrt{s_{\mathrm{thr}}}=2m_p$, 
up to the full $\epem$ CM energy, $\sqrt{s_{\epem}}$. 
Similar formalism can be applied for the annihilation $\epem\to\bar n n$.

In Ref. \cite{Lees:2013xe}, based on the work of Ref.
\cite{Bonneau:1971mk}, the differential cross section for the radiative
process, integrated over the nucleon momenta, was factorized into a
function which depends on the photon kinematical variables multiplied by
the annihilation cross section of interest, for the process 
$\epem \rightarrow \pp$: 
\be 
\displaystyle\frac{d^2\sigma_{\epem \to \pp\gamma}}{ d\sqrt{s_{\epem}} \,d\cos(\thg) }=
\displaystyle\frac{2\sqrt{s}}{s_{\epem}} W(s_{\epem},E_\gamma,\thg
)\sigma_{\epem \to \pp }(s)\,,\hspace{10mm}E_\gamma=\frac{s-s_{\epem}}{2\sqrt{s_{\epem}} }
\,, 
\label{eq:eqisr} 
\ee
where $\sqrt{s}$ and $\sqrt{s_{\epem}}$ are the
invariant masses of the $\pp$ and $\epem$ systems, $E_\gamma$ and
$\thg$ are the  energy and the scattering angle of the photon in the
$e^+e^-$ CM frame, while $W(s_{\epem},E_\gamma,\thg )$ represents the
so-called radiator function, it gives the probability that an initial
photon with energy $E_{\gamma}$ is emitted at the angle $\theta_\gamma$.
In Eq.~\eqref{eq:eqisr}, the factorization of the photon variables
allows to single out the elementary cross section $ \sigma_{\epem \to
\pp}$ and extract the moduli of the TL proton FFs. However, such a
factorization does fail in describing the scattering process when
$\sin(\thg)\to 0$, $i.e.,$ when the photon is radiated along the beam
direction, because it neglects terms depending on $(m_e^2/s_{\epem})$,
where $m_e$ is the electron mass, which become important at small
angles~\cite{Benayoun:1999hm,Baier:1973ms}. The case of final state
radiation (FSR), when the radiative emission is from the final proton or
anti-proton, was discussed in Ref.~\cite{Bytev:2011pa}. It has been
found that also the ISR-FSR interference may spoil the factorization
hypothesis, if the detection is not symmetric around the colliding beams
axis.

The differential cross section for the annihilation process $ e^++e^-\to
\bar p+p $ in Born approximation and in the CM frame
is~\cite{Zichichi:1962ni} 
\be
\frac{d\sigma_{\epem\to\pp}}{d\Omega}(s,\theta)= \frac
{\alpha^2\beta\,{\cal C}(\beta)}{4s}\left
[\left(1+\cos^2(\theta)\right)|G_M(s)|^2+\displaystyle\frac{1}{\tau} 
\sin^2(\theta)|G_E(s)|^2\right ]\,, 
\label{eq:diffcs} 
\ee where $s$ is
the total energy squared of the $\pp$ system, $\tau=s/(4m_p^2)$ and $\beta=\sqrt{1-1/\tau}$ is
the final particle velocity. The function 
$$ 
{\cal
C}(\beta)=\frac{y(\beta)}{1-e^{-y(\beta)}}\,,\hspace{10mm}
y(\beta)=\frac{\pi\alpha}{\beta}\sqrt{1-\beta^2}\,, 
$$ 
represents the Coulomb correction that accounts for the \pp\ final state
interaction~\cite{Hoang:1996nf}.  It becomes effective ($\gg 1$) and
divergent as $\beta\to 0$. Such a divergency, that happens exactly at
the production threshold, $i.e.$, at $\beta=0$ or equivalently at
$s=4m_p^2$, does cancel out the phase-space factor $\beta$ by making
finite and different from zero the cross section at the threshold.

The even $\cos\theta$-angular dependence of the cross section of Eq.~\eqref{eq:diffcs}, in particular the
presence of the powers zero and two only,  results directly from the Born
approximation, $i.e.,$ from the assumption of one-photon exchange and
the invariance of the electromagnetic interaction with respect to the
parity transformation.

Following Ref.~\cite{TomasiGustafsson:2001za}, in order to 
highlight the angular dependence, the  Born differential cross section given in
Eq.~\eqref{eq:diffcs} can be written as 
\be
\frac{d\sigma_{\epem\to\bar p p}}{d\Omega}(s,\theta)=
\sigma_0(s)\left [ 1+{\cal A}(s) \cos^2(\theta)
\right],  \ \sigma_0(s) =\frac{\alpha^2\beta\,{\cal C}(\beta)}{4s}\left(|G_M(s)|^2+
\frac{1}{\tau}|G_E(s)|^2 \right)\,,
\nonumber
\ee 
where $\sigma_0(s)$ is the differential cross section
at $\theta=\pi/2$, and the function ${\cal A}(s)$, assuming the one-photon
exchange mechanism, depends on the ratio of the FFs moduli $R(s)=|G_E(s)|/|G_M(s)|$, as
\be 
{\cal A}(s)=\displaystyle\frac{\tau|G_M(s)|^2-|G_E(s)|^2}
{\tau|G_M(s)|^2+|G_E(s)|^2}=
\displaystyle\frac{\tau - R(s)^2}{\tau+ R(s)^2} \,. 
\label{eq:asym} 
\ee 
It follows that ${\cal A}(s)$ represents an observable which is 
sensitive to deviations of the differential cross section from linearity in
$\cos^2(\theta)$, in particular, a residual dependence on the scattering
angle $\theta$, $i.e.,$ a non null derivative $d\mathcal{A}/d\theta$,
would mean that, besides the one-photon exchange, other intermediate states do
contribute to the annihilation process $\epem\leftrightarrow\pp$.
Similar studies can be made for the scattering processes $e^-p\to e^-p$, in the 
SL region, by considering the deviation from linearity of the so-called 
Rosenbluth plots, see Ref.~\cite{Gakh:2005wa} and references therein.

The total cross section $\sigma_{\epem\to\pp}(s)$, obtained by
integrating  the differential cross
section given in Eq.~\eqref{eq:diffcs} over the solid angle $d\Omega$, namely
\be 
\sigma_{\epem\to\pp}(s)=
\frac{4\pi\alpha^2\beta \,{\cal C}(\beta) }{3s} \left(|G_M(s)|^2+
\frac{1}{2\tau} |G_E(s)|^2\right) \,,
\label{eq:total} 
\ee
is proportional to an $s$-dependent combination
of the moduli squared of the FFs, which is commonly defined in terms of the
effective FF, $F_p(s)$, whose modulus squared is given by the normalized
combination : 
\be 
|F_{p}(s)|^2= \displaystyle\frac{
2\tau|G_M(s)|^2+|G_E(s)|^2}{2\tau+1}\,. 
\label{eq:Geff} 
\ee 
Using such a
unique effective FF is equivalent to consider the protons as a spin-zero
particle and hence, to assume $|G_E(s)|=|G_M(s)|\equiv |F_p(s)|$ in
Eq.~\eqref{eq:diffcs}. As a consequence of their definitions in terms of
the Dirac and Pauli FFs, $F_1(s)$ and $F_2(s)$ :  
\be
G_E(s)=F_1(s)+\tau\,F_2(s)\,,\hspace{10mm}
G_M(s)=F_1(s)+F_2(s)\,,\hspace{10mm} 
\nonumber 
\ee 
and the assumption of analyticity, the identity $G_E(s)=G_M(s)$ is strictly valid only at the
production threshold $s=4m_p^2$, $i.e.$, $\tau=1$. This
phenomenon can be also interpreted as a consequence of the isotropy of
the annihilation process $\epem\to\pp$ just at the production threshold,
in the \pp\ or \epem \ CM frame. In fact, having no preferred direction,
the amplitude must be independent on the scattering angle, that implies
$\mathcal{A}(4m_p^2)=0$, see Eq.~\eqref{eq:asym}, $i.e.,$
$G_E(4m_p^2)=G_M(4m_p^2)$. %

Therefore, a measurement of the total cross section gives access to the
effective FF.  The extraction of $R$  and/or $\mathcal{A}$ requires in
addition a precise measurement of the differential cross section.  Even
further precision is required for a meaningful extraction of the
individual FFs. It is for this reason that it could be achieved only in
the most recent experiments.  \section{Analysis
of the results}
\subsection{Selected data sets} 

We consider four sets of data on the $\sigma_{\epem\to\pp}$ cross
section.

\begin{enumerate}

\item  The set from the BaBar Collaboration, labeled as ``BaBar'', has
three sub-sets:

\begin{itemize} 
\item 38 points, obtained with the ISR technique and
detecting the initial photon, in the range  $(1.877  \le\sqrt{s} \le
4.50)$~GeV, together with 6 points for the ratio $R=|G_E|/|G_M|$ in the
range $(1.877\le\sqrt{s} \le3)$~GeV~\cite{Lees:2013xe}. 
\item  13 points, obtained with the ISR technique and detecting the initial
photon, in the range $(1.8765 \le\sqrt{s} \le 1.9625
)$~GeV~\cite{Lees:2013xe}. These data with a larger granularity overlap
with the first four points of the above series, that, therefore, are
omitted in the analysis. 
\item  8 points, obtained with the ISR
technique and not detecting the initial photon, in the range
$(3\le\sqrt{s}\le5.50)$~GeV~\cite{Lees:2013uta}. 
\end{itemize}
\item  Two sets from the BESIII Collaboration:

\begin{itemize} \item 30 points, obtained with the ISR technique, in the
range $(2.0 \le\sqrt{s}\le 3.60)$~GeV~\cite{Ablikim:2019njl}, labeled as
``BESIII-ISR''. \item 22 points, obtained with energy scan, together
with 16 points for the ratio $R$, and for the disentangled moduli
$|G_E|$ and $|G_M|$, in the range $(2.0 \le\sqrt{s} \le 
3.08)$~GeV~\cite{Ablikim:2019eau}, labeled as ``BESIII-BS''.
\end{itemize}

\item    A set from CMD-3 of 11 points, obtained with energy scan, in the
range $2m_p< \sqrt{s}\le 2.006)$~GeV ~\cite{CMD-3:2018kql}. They belong
to a sub-set of the published data, that includes only those points
lying above the production threshold $\sqrt{s}=2m_p$. Indeed, the
complete set covers an energy interval that, as a consequence of
experimental limits of the energy resolution, extends also below the
physical threshold. This is the second measurement of the
$\sigma_{\epem\to\pp}$ cross section performed by the CMD-3 Collaboration
and it improves the first one~\cite{Akhmetshin:2015ifg} by enhancing the
precision and extending the energy range. As numbers are not given in
the original paper, the points (red squares in Fig. 4 of Ref.
\cite{CMD-3:2018kql}) have been read from the figure. 
This set of data is labeled as CMD-3.

\end{enumerate}
\subsection{Confirmation of the oscillations}

In Ref. \cite{Bianconi:2015owa} it was pointed out that the cross
section  of $\epem\to\pp$ measured by the BaBar Collaboration
\cite{Lees:2013xe}  shows evidence of structures. These structures
become regular when plotted  as a function the 3-momentum $p$ of one of
the two hadrons in the frame where the other one is at rest and it is
proportional to the relative velocity $\beta=\sqrt{1-1/\tau}$.

The BaBar data on the modulus of the proton effective
FF~\cite{Lees:2013xe}, extracted from the $\epem\to\pp$ total cross
section by means of the formulae given in Eq.~\eqref{eq:total}
and~\eqref{eq:Geff}, in the range $(2m_p<\sqrt{s}<6)$~GeV, are well
reproduced by the function  \cite{Bianconi:2015owa} 
\be F_p^{\rm fit}(s)=F_{\rm 3p}(s)\ +\ F_{\rm osc}(p(s))\,. 
\label{eq:diff} 
\ee 
It is the sum of two contributions: a dominant  three-pole (3p) 
$F_{\rm 3p}(s)$, and a damped oscillatory component 
$F_{\rm osc}(p(s))$, whose
expressions  are 
\begin{eqnarray} 
F_{\rm 3p}(s)&=& \frac{F_{0}} {\left(
1+\frac{s}{m_a^2}\right)\left(1-\frac{s}{m_0^2} \right)^2}
\label{eq:f3p} \,,\\ 
F_{\rm osc}(p(s))&=&Ae^{-Bp}\cos(Cp+D)\,.
\label{eq:fosc} 
\end{eqnarray} 
The explicit expressions of the variables
$p=s(p)$ and $p=p(s)$,  as well as of the functions in terms of $s$ and
$p$ are explicited in the Appendix.

The 3p function $F_{\rm 3p}(s)$, that describes the smooth behavior
(ignoring small-scale oscillations) of the effective FF, is the product
of a free monopole, depending on two free parameters: the adimensional
$F_0$ and the mass $m_a$, and the standard dipole with $m_0^2=0.71$
GeV$^2$.

The oscillatory contribution $F_{\rm osc}(p(s))$, reproduces the
GeV-scale oscillations in the $p$ variable. These irregularities are
treated as small perturbations of the dominant smooth behavior, $i.e.,$
$|F_{\rm osc}(p(s))|\ \ll \ |F_{\rm 3p}(s)|$. Moreover, due to their
regular periodic nature, they have a vanishing mean effect \be \langle 
F_{\rm osc}(p(s)) \rangle_{\Delta p}\mathop{\longrightarrow}_{\Delta
p\geq 1\,{\rm GeV}} 0 \,.\nonumber \ee

Here we  show that the recent data on $F_{P}$ from the BESIII 
Collaboration  \cite{Ablikim:2019njl, Ablikim:2019eau} are %
compatible with those from the BaBar Collaboration 
\cite{Lees:2013xe, Lees:2013uta} and confirm the previous findings 
of Ref.  \cite{Bianconi:2015owa}.   This is proved by the 
consistency of the fit parameters,  obtained by including the data 
sets BESIII-ISR and BESIII-SC, besides the BaBar one, compared to 
the parameters obtained by fitting the BaBar data only (Table~\ref{tab:fit1}).

\begin{table} 
\caption{\label{tab:fit1} Fit parameters  from Eq.~\eqref{eq:f3p}  } 
\begin{ruledtabular} 
\begin{tabular}{c|c|c|c|c}
 Ref. & Exp. &N        & $ F_0$ & $m_a^2$ (GeV$^2$)\\ \hline
\protect\cite{Lees:2013xe, Lees:2013uta, Bianconi:2015owa}&  BaBar  & 85
& 7.7 $\pm$ 0.3 & 15 $ \pm$ 1 \\
\protect\cite{Ablikim:2019njl,Ablikim:2019eau,Lees:2013xe,Lees:2013uta}
&BaBar,BESIII-ISR,BESIII-SC & 107     & $ 8.9 \pm 0.2$ & $8.8\pm 0.6$\\ 
\end{tabular} 
\end{ruledtabular} 
\end{table}

\begin{figure} 
\begin{center}
\includegraphics[width=8.3cm]{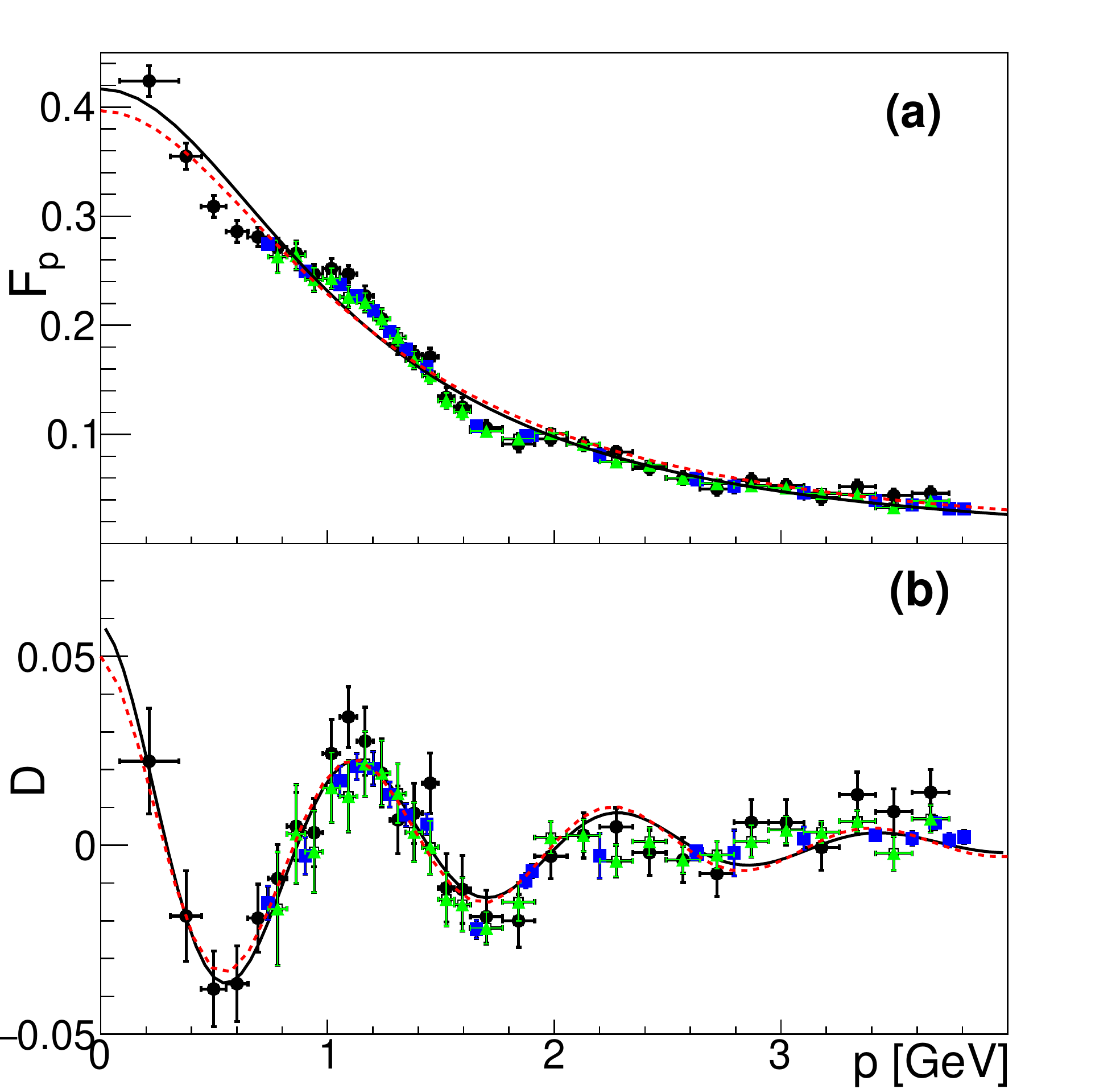} 
\caption{(a): TL
proton generalized FF as a function of $p$ from the data of BaBar, Ref.
\protect\cite{Lees:2013xe} (black circles), BESIII-ISR
\protect\cite{Ablikim:2019njl}  (blue squares) and  BESIII-SC
\protect\cite{Ablikim:2019eau} (green triangles), with the regular
background fit with Eq. \eqref{eq:f3p}  (black solid line );  (b): data
after subtraction, fitted with Eq. \eqref{eq:fosc} (black solid line).
For comparison the fit from Ref. \cite{Bianconi:2015owa} (red dashed
lines) is also shown. } \label{Fig:FitDiff} 
\end{center} 
\end{figure}

In Fig. \ref{Fig:FitDiff}a  the cross section data are plotted as a
function of $p$. The result of the fit using Eq.~\eqref{eq:f3p} is then
subtracted from the data. The obtained residue ${\cal D}$ (data
minus $F_{\rm 3p}(s)$) displayed in Fig. \ref{Fig:FitDiff} shows a
damped and periodic oscillatory behavior, that has been fitted with the
four-parameter function of Eq.~\eqref{eq:fosc}. The values  of the parameters are reported in Tables
\ref{tab:fit1}, \ref{tab:GEff},  together with those obtained by fitting the BaBar
data only. 
\begin{table} 
\caption{\label{tab:GEff} 
Fit parameters  from Eq.
\eqref{eq:fosc} and corresponding values of the normalized $\chi^2$. }
\begin{ruledtabular} 
\begin{tabular}{c|c|c|c|c|c|c} 
Ref. & Data set & $
A \pm \Delta A $ & $B \pm \Delta B$ & $C \pm \Delta C$ & $D \pm \Delta
D$ & $ \chi^2/n.d.f$   \\
 &&  & (GeV$^{-1}$)    &
(GeV$^{-1})$ & &\\ 
\hline
\protect\cite{Lees:2013xe,Lees:2013uta,Bianconi:2015owa}&  BaBar & $0.05
\pm   0.01$ &$ 0.59 \pm  0.2$ &$5.6 \pm  0.1 $ &$0.2 \pm 0.2$
&57/(55-4)= 1.1      \\ 
\hline
\cite{Ablikim:2019njl,Ablikim:2019eau,Lees:2013xe,Lees:2013uta}
&BESIII-ISR,SC,BaBar  & $0.07\pm 0.01$ &$ 0.93 \pm  0.09$ &$5.9 \pm 0.1
$ &$0.1 \pm 0.2$ &227/(107-4)=2.2 \\ 
\end{tabular} 
\end{ruledtabular}
\end{table} 

As shown in Fig.~\ref{Fig:FitDiff}, even with a slightly worse
normalized $\chi^2$, the new fit (black solid line) follows closely the
one on the only BaBar data~\cite{Bianconi:2015owa} (red dashed line).
Let us note that the consistency of the data obtained with different
methods, beam scan and ISR, rules out the possibility that the
oscillations could be an artefact of the ISR technique or of the photon
detection.
\section{Global Fit of the data }

\begin{figure} 
\begin{center}
\includegraphics[width=12cm]{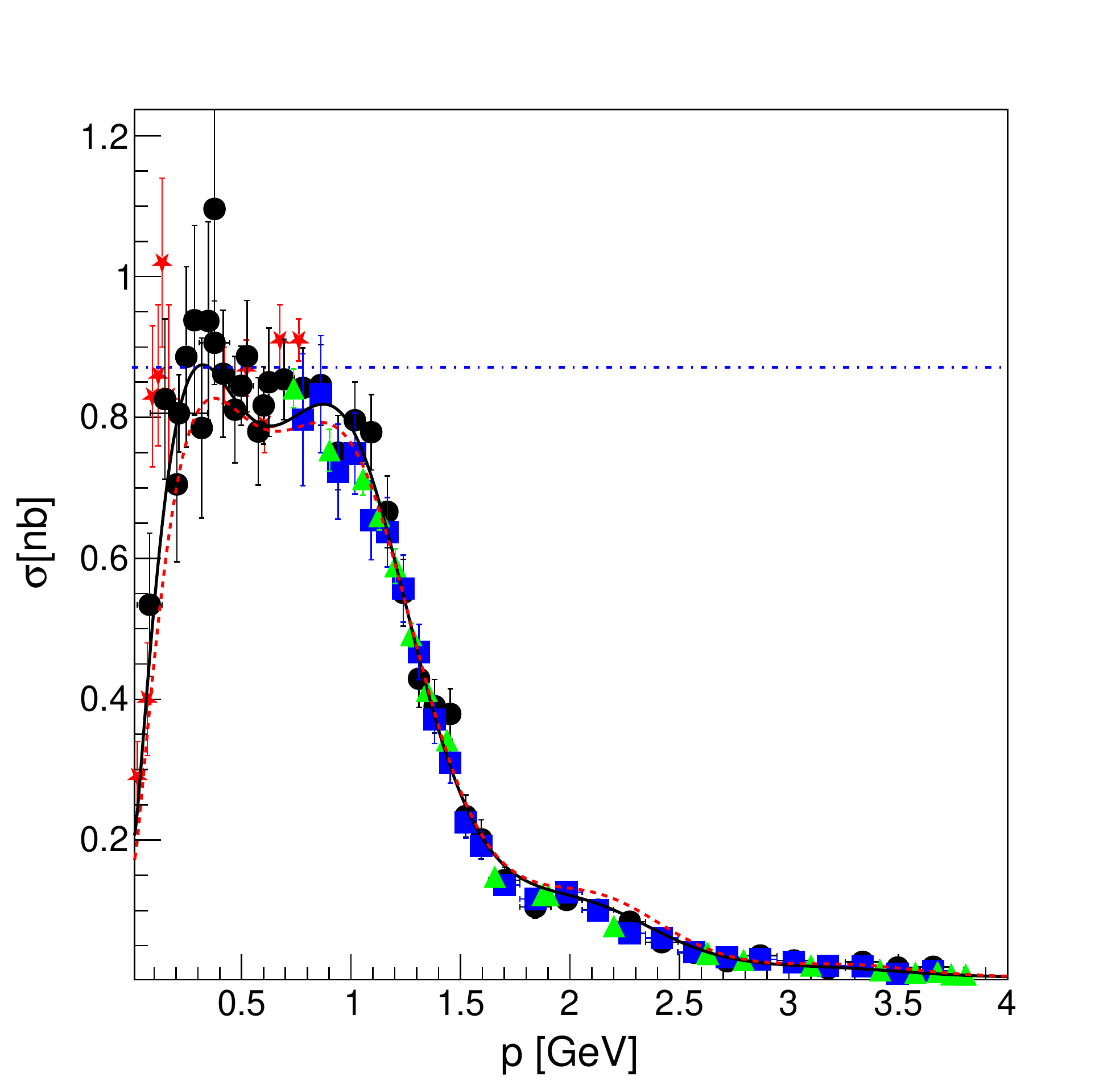} 
\caption{Born cross
section for $e^++e^-\to p + \bar p$ as a function of the momentum $p$.
The data  are from CMD-3 \cite{CMD-3:2018kql} (red stars), BaBar
\protect\cite{Lees:2013xe,Lees:2013uta} (black circles) BESIII-ISR
\protect\cite{Ablikim:2019njl} (blue squares)  and BESIII-SC
\protect\cite{Ablikim:2019eau} (green triangles) are shown together with
the six-parameter fit from Eqs.
(\ref{eq:diff},\protect\ref{eq:f3p},\ref{eq:fosc}) and Table
\ref{Table:6par}(black solid line), compared to the fit from Ref.
\protect\cite{Bianconi:2015owa} (red dashed line). The blue dash-dotted
line corresponds to a constant, fitted in the range $0.1 <\sqrt{s}< 0.9 $ GeV.
} 
\label{Fig:6parCS} 
\end{center} 
\end{figure}

The cross section or the effective FF data  can also be directly fitted with the
six-parameter function $F_p^{\rm fit}(p)$ of Eq.~\eqref{eq:diff}.  The parameters are reported in
Table~\ref{Table:6par} and the fit is illustrated in
Fig.~\ref{Fig:6parCS} as a function of the relative momentum $p$ (black
solid line), together with the result from Ref.~\cite{Bianconi:2015owa}.

\begin{table}[h] 
\caption{\label{Table:6par} 
Six-Parameters fit,
Eq.~\eqref{eq:diff}, of the annihilation cross section
$\sigma_{\epem\to\pp}$ as a function of relative momentum $p$ for the
BaBar, BESIII and CMD-3 data.} \small \renewcommand{\arraystretch}{1.3}
\small\addtolength{\tabcolsep}{1 pt} \begin{tabular}{c|c|c|c|c|c|c|c}
\hline \hline \multirow{2}{*}{Ref.} &\multirow{2}{*}{$F_0$} &$m_a^2$   &
\multirow{2}{*}{$ A $} & $B $ & $C $ & \multirow{2}{*}{$D $} &
\multirow{2}{*}{$\displaystyle\frac{\chi^2}{\rm n.d.f.}$ }   \\ 
& & (GeV$^{2}$)  & & (GeV$^{-1}$) & (GeV$^{-1}$) & &  \\ 
\hline
\cite{Ablikim:2019njl,Ablikim:2019eau,Lees:2013xe,Lees:2013uta,CMD-3:2018kql} &   9.7 $\pm$ 0.3  &  7.1 $\pm$  0.5   & 0.073 $\pm$ 0.007 &
1.05 $\pm$ 0.07  &5.51$ \pm$ 0.09   &0.04 $\pm$ 0.1       &
$\displaystyle \frac{278}{118-6}=    2.5$      \\ 
\hline \hline
\end{tabular} 
\end{table}

Extending the data sets does not change essentially the fit, 
worsening the $\chi^2$. The inclusion of the CMD-3 data heightens the curve
in the near threshold region. The blue dash-dotted line
corresponds to a constant fitted in the range $0.1<\sqrt{s}
<0.9$~GeV, that gives the average value of the cross section 
$\bar\sigma= 0.87\pm 0.02$ nb. Such a value is close to the cross 
section at the production threshold for a structureless 
fermions~\cite{Baldini:2007qg}, as for instance that of the 
reaction $\epem\to\mu^+\mu^-$.

\begin{figure} 
\begin{center}
\includegraphics[width=12cm]{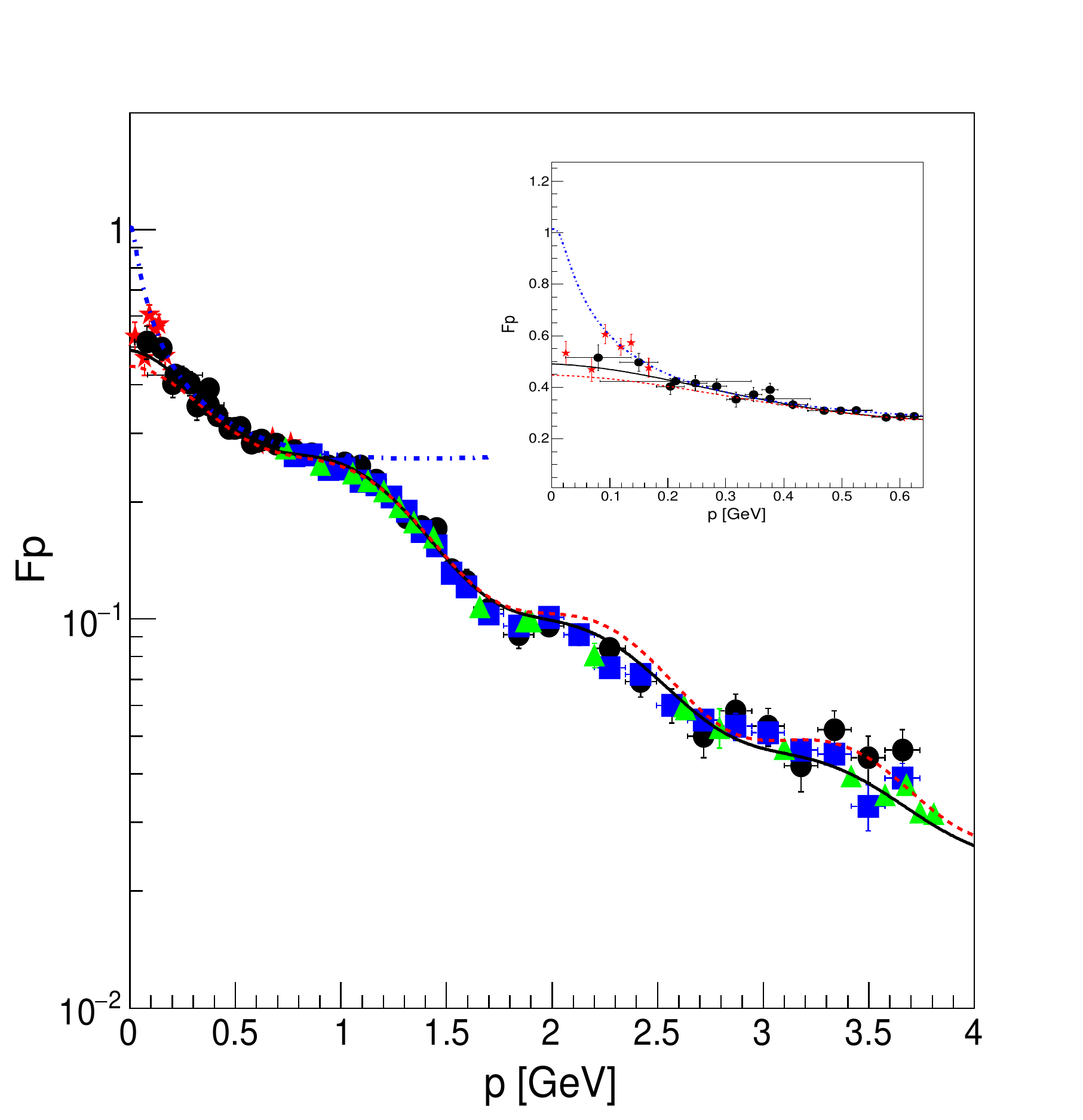} 
\caption{Same as Fig.
\ref{Fig:6parCS} but for the TL proton generalized FF. The blue
dash-dotted line is the expectation for a constant cross section
$\sigma=0.87$ nb. } 
\label{Fig:6parFP} 
\end{center} 
\end{figure}

Fig.~\ref{Fig:6parFP} shows the data on the effective FF together with
the curves that represent the corresponding fit functions.

\subsection{Analysis of the form factor ratio $R$}

The comparison between the absolute values of the electric and magnetic
FFs in  the TL and SL regions can be more easily done by
considering their ratio. Exploiting the Akhiezer-Rekalo recoil proton
polarization method~\cite{Akhiezer:1968ek,Akhiezer:1974em}, that
represents a unique and very powerful technique to extract directly the
FF ratio $G_E/G_M$ from the longitudinal to transverse recoil proton
polarization in the elastic scattering process ${\vec e}^{\,-} p \to e^-
\vec p$, the JLab-GEP Collaboration obtained very precise values of $R$
in a wide region of transferred momenta~\cite{Puckett:2017flj,Puckett:2011xg}. Note
that the individual FFs can not be determined by this method. Therefore
it is assumed that the magnetic FFs is well known from the unpolarized
cross section measurements.

\begin{figure} 
\begin{center}
\includegraphics[width=8.cm]{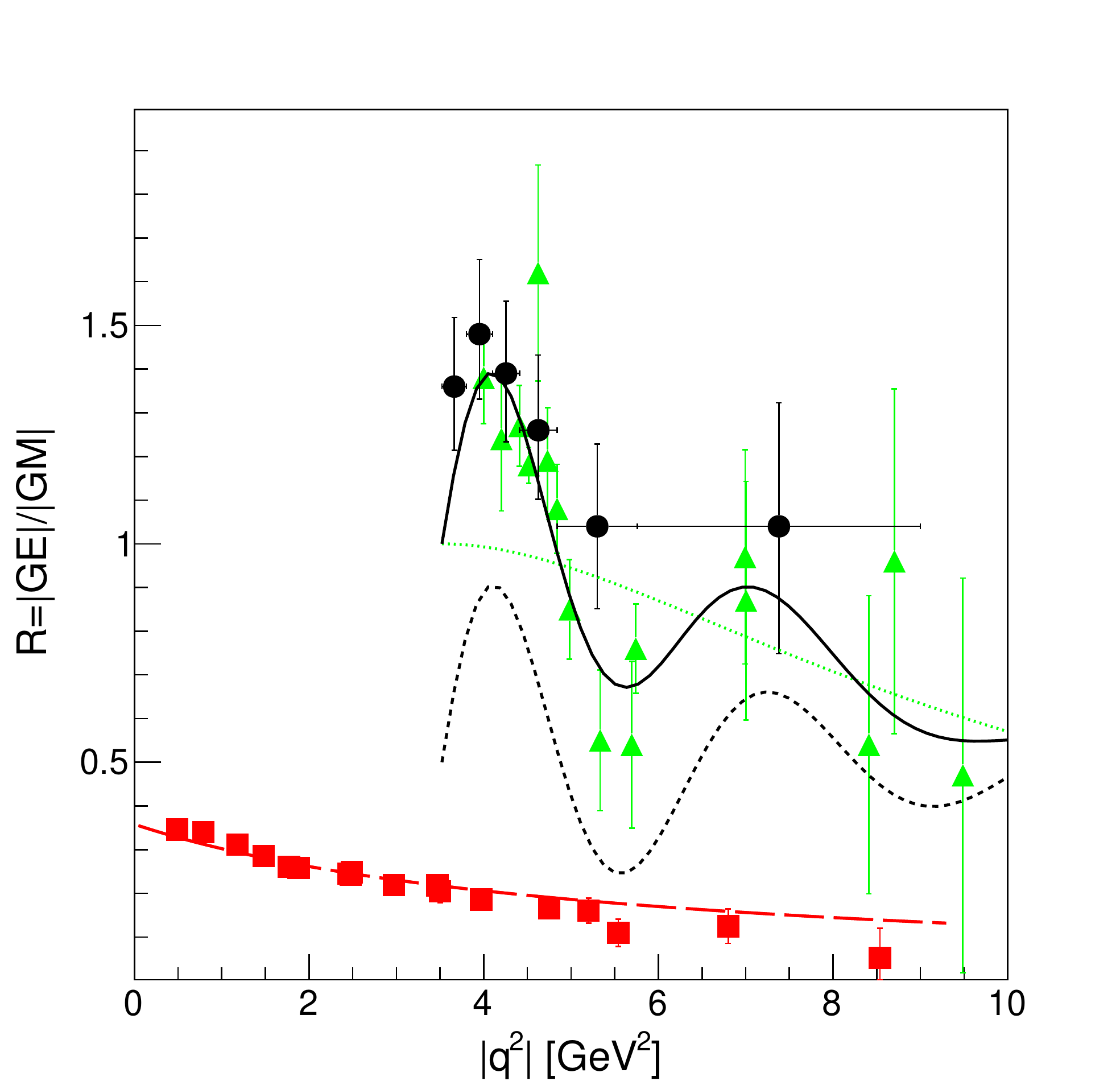}
\includegraphics[width=8.cm]{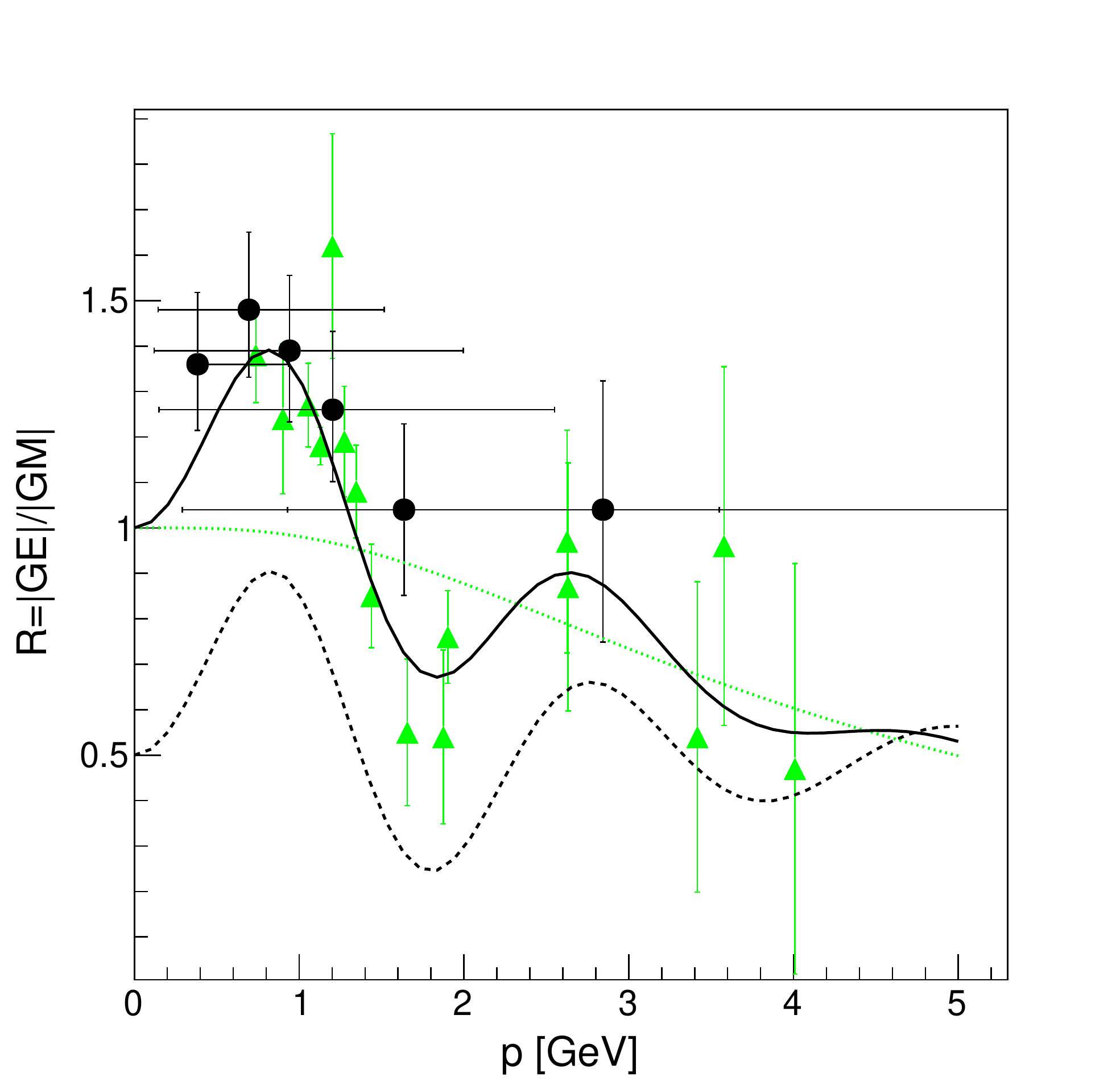} 
\caption{Ratio
$R=|G_E|/|G_M| $ as a function of $|q^2|$ (left) and $p$ (right) from
BaBar  (black circles) \protect\cite{Lees:2013xe,Lees:2013uta} and  BES-SC 
(green triangles) \cite{Ablikim:2019eau}. The  solid black line  is the
fit from Eq.~(\ref{eq:FR}), decomposed in the monopole component (green
dashed line) and the oscillatory component (black dotted line - shifted
up by 0.5). The SL ratio from the JLab-GEp Collaboration 
\protect\cite{Puckett:2017flj} is also shown (red squares), together
with its constrained monopole fit (red long-dashed line). }
\label{Fig:RatioTLSL} 
\end{center} 
\end{figure}

In the TL region, the present data from the BESIII Collaboration
bring new information on the ratio of the FFs moduli with comparable
precision as in the scattering region. The data from BaBar and BESIII
are plotted in Fig. \ref{Fig:RatioTLSL} as a function of $|q^2|$. The
choice of this variable does allow to show on the same graph SL and TL
values of the FF ratio and of  their moduli
respectively~\cite{TomasiGustafsson:2001za}.

While the SL data (red squares in Fig.~\ref{Fig:RatioTLSL}) show
a monotone decrease, the TL ones (green triangles in
Fig.~\ref{Fig:RatioTLSL}) ~\cite{Ablikim:2019eau}, decrease
too, but show the presence of oscillations, not contradicting the
results from BaBar~\cite{Lees:2013xe} (black circles in
Fig.~\ref{Fig:RatioTLSL}). One can see a minimum in the TL range
$(5-6)$~GeV$^2$, in correspondance to a little dip in the SL 
region, that should be  confirmed, because it lies just at the square
momentum transfer corresponding to the kinematical limits of two
experiments  of the JLab-GEP Collaboration.

The SL and TL values of the FF ratio, 
move away with a smooth decrease 
from $1/\mu_p$ ($\mu_p$ is the proton magnetic moment in units
of the Bohr magneton) at $q^2=0$,  and from unity at the production
threshold $q^2=4m_p^2$, respectively. These are the values expected from the definitions given above, as well as, at large transferred momenta, 
from the QCD  quark counting rules
~\cite{Matveev:1973uz,Brodsky:1973kr}. This is an indication that the
perturbative domain has not been reached and corroborates the
predictions from Ref. \cite{Kuraev:2011vq}. Following a similar approach
as for the effective FF, we fit the ratio in the TL region with a function $F_R$
reproducing a monopole decrease and a damped oscillation: 
\be 
F_R(\omega(s))=
\displaystyle\frac{1}{1+\omega^2 /r_0} \left [1+r_1e^{-r_2 \omega }\sin\left(r_3 \omega
\right)\right ], \  \omega=\sqrt{s}-2m_p \,, 
\label{eq:FR} 
\ee 
where the unitary normalization at the production threshold, $F_R(4m_p^2)=1$, is
imposed. The curve representing the fit function, Eq.~\eqref{eq:FR},
obtained with the parameters reported in Table~\ref{Tab:Rfit}, is shown
as a black line in Fig.~\ref{Fig:RatioTLSL}, together with the
corresponding data on the TL ratio $R$ (black circles and green
triangles). The monopole and the oscillatory components are also shown. 
\begin{table}[h]
 \caption{\label{Tab:Rfit} 4-parameters fit for $R$ as a
function of $q^2$. } 
\renewcommand{\arraystretch}{1.3}
\small\addtolength{\tabcolsep}{5pt} 
\begin{tabular}{c|c|c|c|c} 
\hline \hline 
$ r_0$   &  \multirow{2}{*}{$ r_1$} & $r_2$  & $r_3 $ &
\multirow{2}{*}{$\displaystyle\frac{\chi^{2}}{\rm n.d.f }$} \\
(GeV$^{2}$) &  & (GeV$^{-1}$) &  (GeV$^{-1}$) & \\ \hline 3 $\pm$ 2 &
0.5 $\pm$  0.1 & 1.5 $\pm$ 1.2 & 9.3 $\pm$ 0.5 &
$\displaystyle\frac{14}{22-4}=0.8$  \\ 
\hline \hline 
\end{tabular}
\end{table}

The red long-dashed line in Fig.~\ref{Fig:RatioTLSL} visualizes a
one-parameter monopole function, constrained to $1/\mu_p$ at $q^2=0$.
Let us remind that in the space-like region the electric FF is
normalized to 1 (in unit of electric charge) and the magnetic FF is
normalized to $\mu_p$ at $q^2=0$ .

In Ref.~\cite{Kuraev:2011vq} it was suggested that a faster decreasing
behavior of the electric FF compared to the magnetic FF in the
SL, as well as in the TL region, is expected as a consequence of
the presence of an inner volume inside the nucleon that is electrically
neutral (short distances corresponding to large transferred momenta).
The consequence is a dipole behavior for the magnetic FF and an
additional monopole decrease for the electric FF, so that  the ratio 
decreases like a monopole.

 \subsection{Zero crossing of the angular asymmetry ${\cal A} $}

\begin{figure} 
\begin{center} 
\includegraphics[width=8.cm]{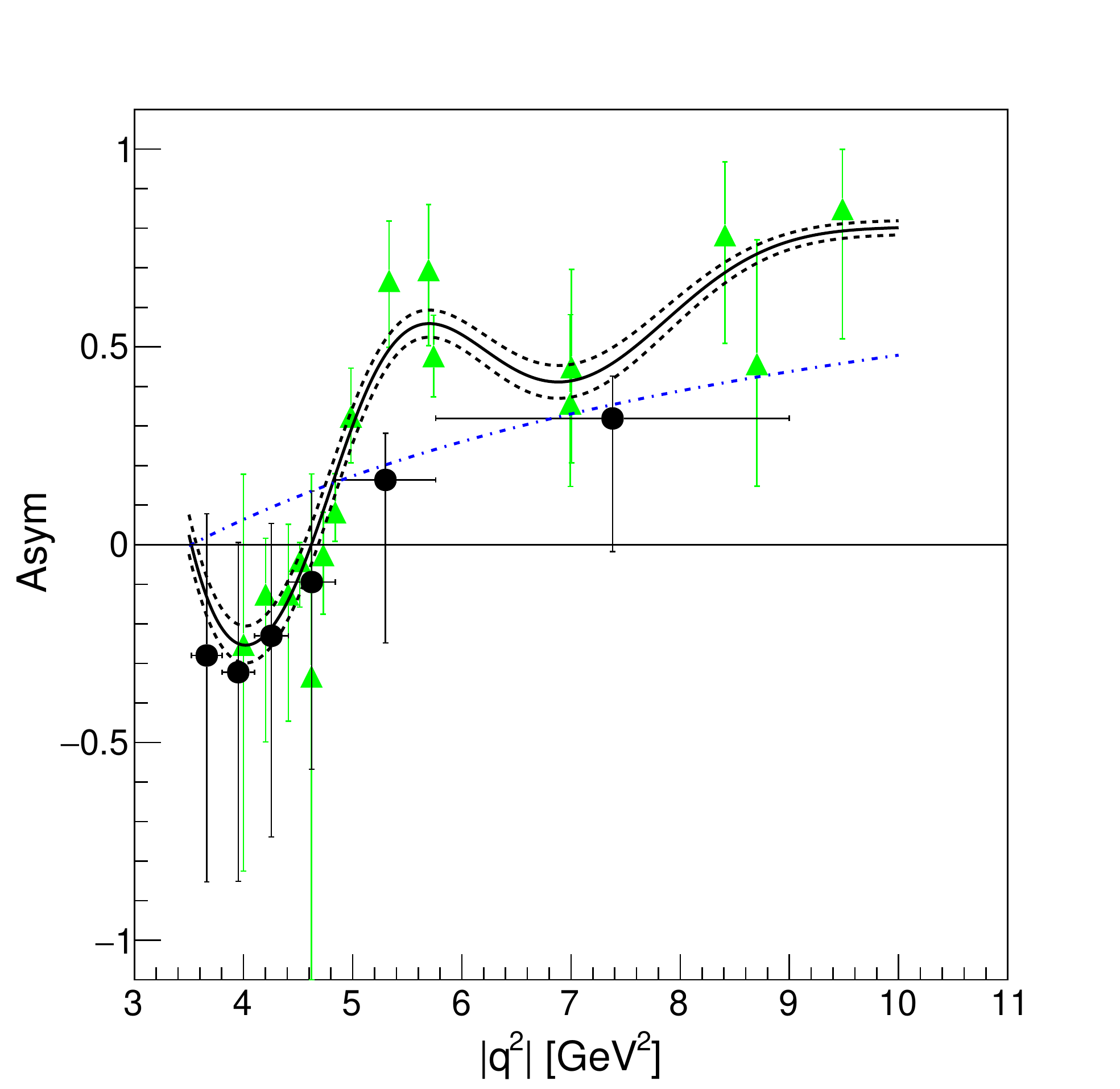}  
\includegraphics[width=8.cm]{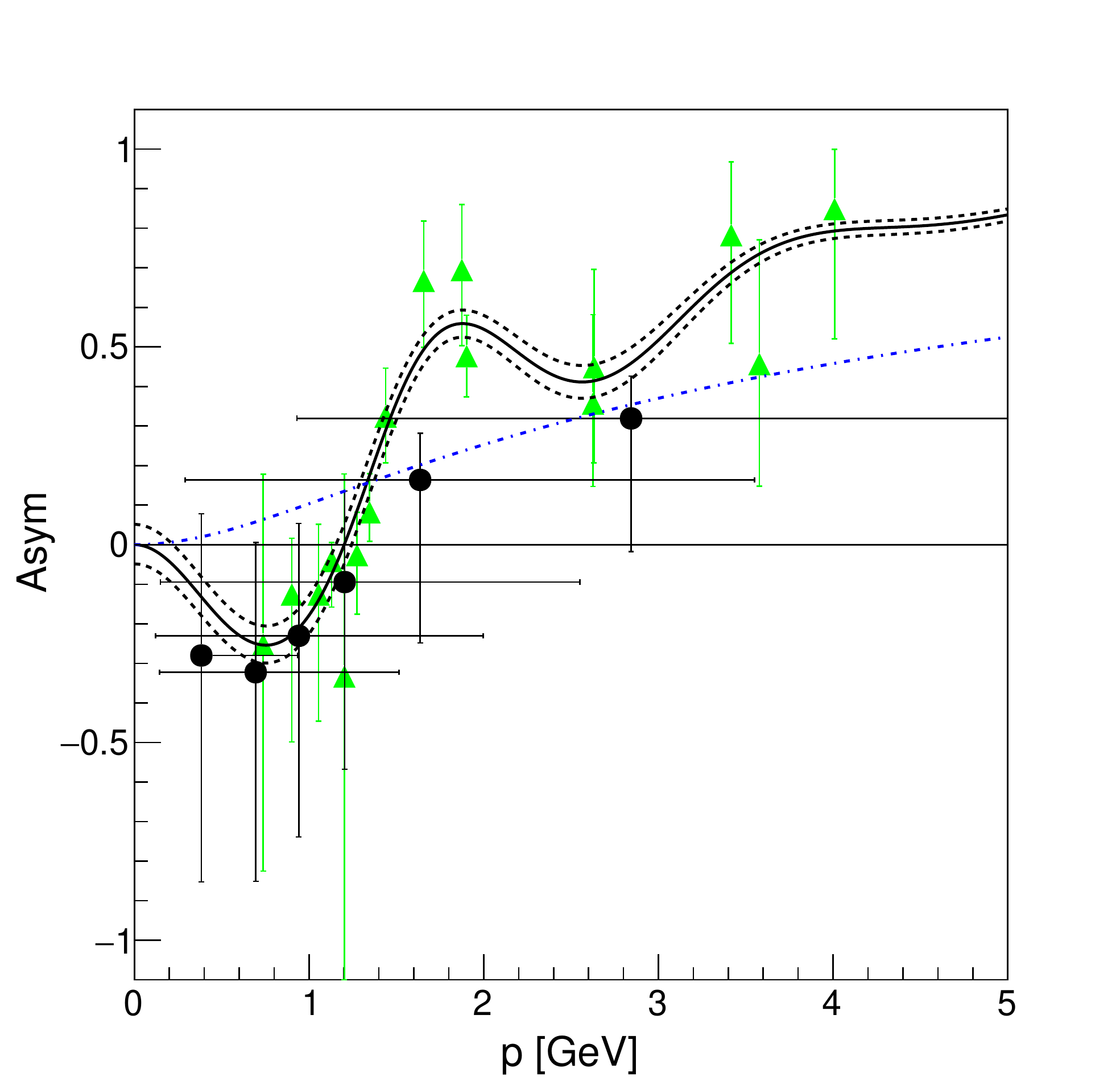} \caption{Angular 
asymmetry as a function of $|q^2|$ (left) and of $p$ (right) 
from BaBar (black circles), BESIII-SC (green triangles). The 
dash-dotted blue curve corresponds to a constant and unitary 
ratio, $ i.e.,$ $|R|=1$, while the solid black curve is related  
to the fit function $F_R(s)$, with the dashed black curves
representing a $\pm 5\%$ variation of the function.} 
\label{Fig:Asym} 
\end{center} 
\end{figure}

A further possibility to illustrate these results, knowing the ratio $R$ and the
fit function, is to calculate the angular asymmetry, ${\cal A}(s)$, 
from  Eq.~\eqref{eq:asym}.  By definition, it assumes values in the
range $[-1,1]$, being null at the production threshold, 
$i.e.,$ ${\cal A}(4m_p^2)=0$.  The data and the fit on 
${\cal A}$ are shown in Fig. \ref{Fig:Asym}.

It has been previously pointed out that, when extracted directly from
the cross section, the relative error on this variable is equivalent to
an error on $R^2$, being therefore preferable for the extraction of the
individual FFs \cite{Singh:2016dtf,Tomasi-Gustafsson:2014pea}.

One can see that ${\cal A}(s)$ crosses zero at $s =(4.62\pm
0.07)$~GeV$^2$, meaning that, also at this squared momentum transferred,
the modulus of the ratio is equal to one, and hence, $|G_E|=|G_M|$. The
uncertainty is obtained by varying the function in a $\pm 5\%$ range
(dashed black lines). The determination of the zero crossing gives a
precise experimental constrain on FF models.

\subsection{Individual form factors $|G_E|$ and $|G_M|$}

The monopole background, used to fit the FF ratio is consistent with
Ref. \cite{Kuraev:2011vq} where it was suggested that the magnetic FF
would follow a dipole dependence, whereas an additional monopole factor
would induce a faster decrease of the electric FF in both SL and
TL regions.

\begin{figure} \begin{center}
\includegraphics[width=8.3cm]{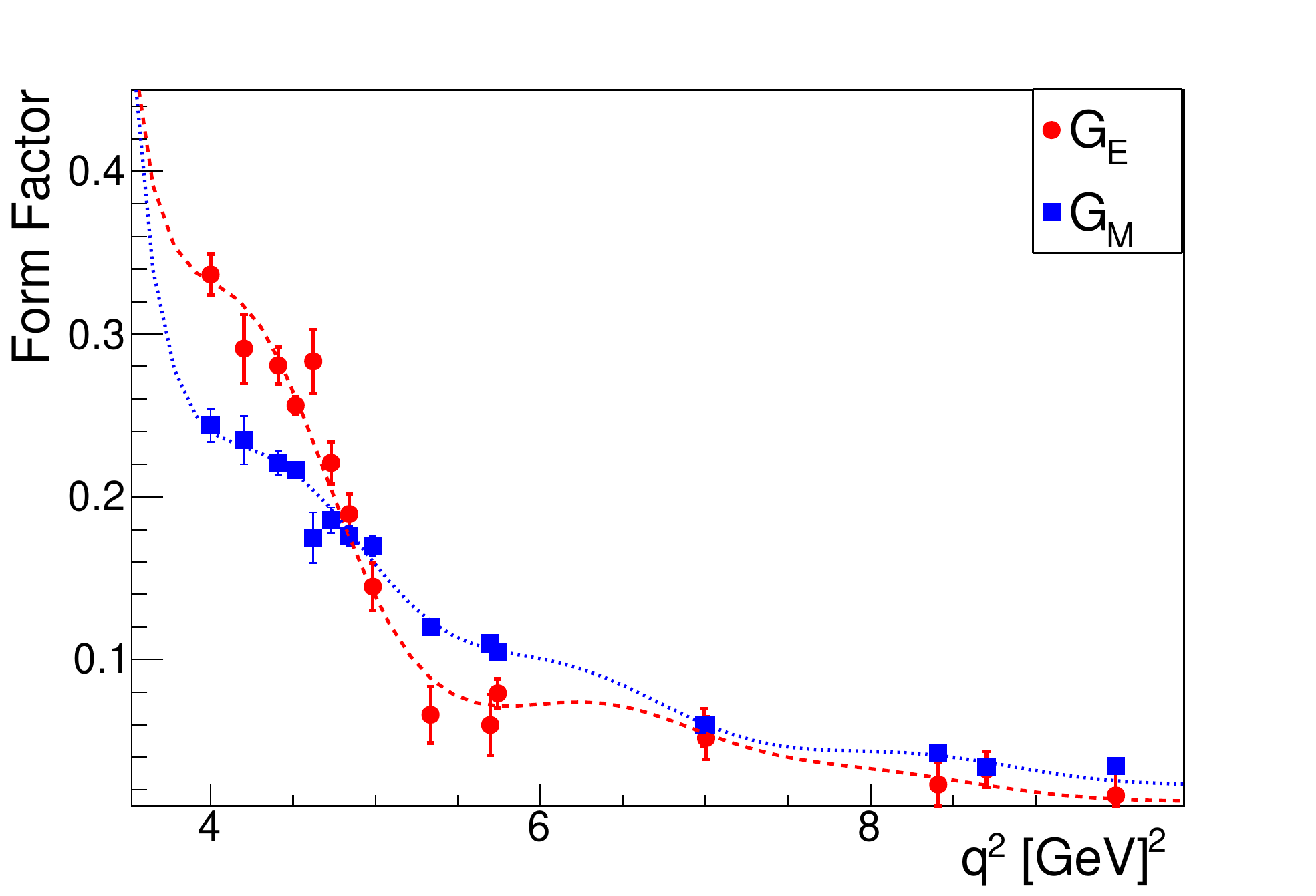} 
\caption{$|G_E| $ (red
circles) and $|G_M|$ (blue squares) from BESIII. 
The dashed red
(dash-dotted blue) line is the calculation  of $|G_E|$ and $|G_M|$ from
the fits of the effective FF $F_p$ and the ratio $R$.} 
\label{Fig:GEGM}
\end{center} 
\end{figure}

Overlapping the data for $|G_E|$ and $|G_M|$, extracted separately for
the first time by the BESIII Collaboration \cite{Ablikim:2019eau}, from
$\epem\to\pp$ differential cross section data, the different behavior of
the two FFs becomes visible and sizeable, as shown in Fig.~\ref{Fig:GEGM}.
Surprisingly, $|G_E| $ and $|G_M| $ are also different at smaller $q^2$,
(even though they should coincide at the production threshold) and 
seem to converge towards small values or to zero at large $q^2$.

One may inquire if the oscillations that are present in the cross
section and in the effective FF are also visible in the individual FFs,
and, in this case,  if they have to be attributed to the electric or the
magnetic FF, or to both of them. The modulus of the electric FF, $|G_E| $, 
shows larger deviations from a
smooth behaviour, in particular it has a dip around 5-6 GeV$^2$, whereas
$|G_M| $ follows closely a $(q^2)^{-2}$ decrease. The relations between the
pairs of functions ($|G_E|$, $|G_M|$) and  ($R$, $F_p$) are: 
\be |G_E(s)|=
F_p(s)\sqrt{ \frac{1+2\tau}{R^2(s)+2\tau/R^2(s)}}\hspace{10 mm} 
|G_M(s)|=
F_p(s)\sqrt{ \displaystyle\frac{1+2\tau}{R^2(s)+2\tau}}. 
\label{eq:gegm}
\ee
By means of these expressions, the moduli of the electric and magnetic
FFs can be calculated using for the ratio $R$ and the effective FF $F_p$
their fit function $F_R$, Eq.~\eqref{eq:FR}, and $F_p^{\rm fit}$, 
Eq.~\eqref{eq:diff}, respectively. The resulting curves are shown in
Fig.~\ref{Fig:GEGM}. This procedure gives by construction a smooth
description of the individual moduli of the two FFs, from threshold up
to the highest experimentally accessible values of $s$ and represents a
particular interest to illustrate the near threshold behavior, 
as the extrapolation of the FF data is constrained by the condition $R(s=4m_p^2)=1$.

The result is shown in Fig. \ref{Fig:GEGM}. Oscillations characterize
both FFs, although they are more smooth on $|G_M|$. By the definition of
the $F_R(s)$ fit function, the convergence of the two electric and
magnetic FFs, and hence, also of the effective one, to a common value at
the production threshold is implied and we find:
$|G_E(4m_p^2)|=|G_M(4m_p^2)|=|F_p(4m_p^2)|\equiv F_{\rm th}\simeq 0.48$.

\begin{figure} 
\begin{center}
\includegraphics[width=15cm]{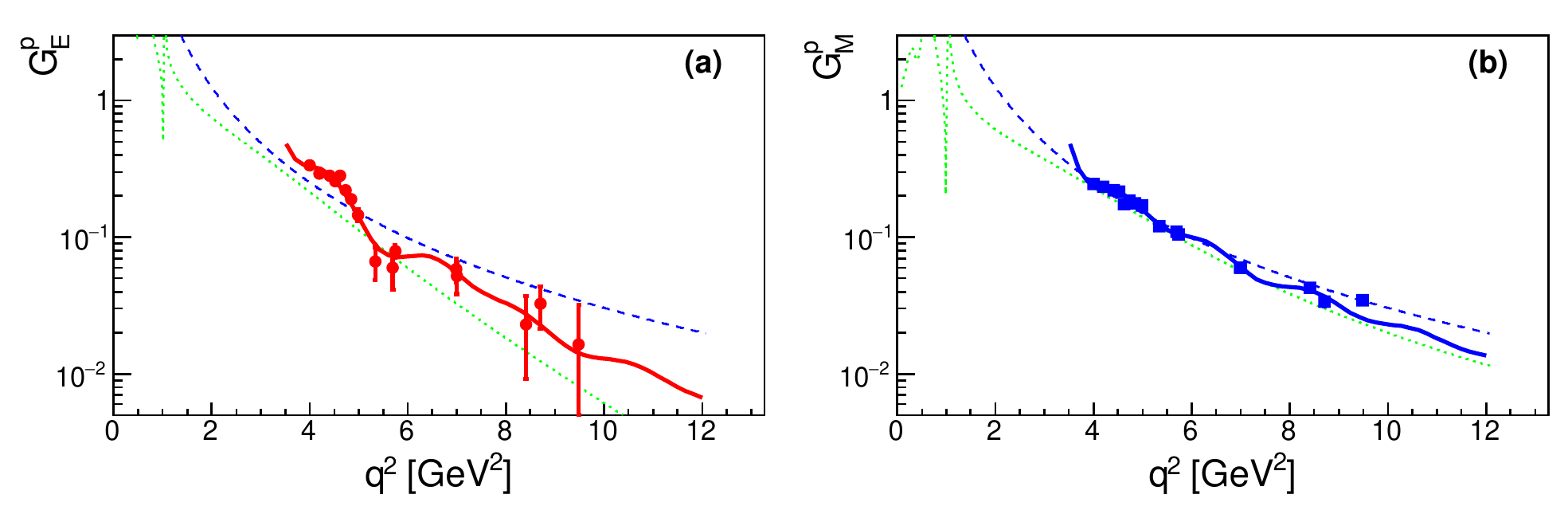} 
\caption{(a) $|G_E| $ (red
circles) and (b) $|G_M|$ (blue squares) from the BESIII Collaboration. The dashed blue (solid
green) curve is the calculation  of $|G_E|$ and $|G_M|$ from the QCD
extrapolation \cite{TomasiGustafsson:2005kc} and from the VMD model of
Ref. \cite{Bijker:2004yu}.} 
\label{Fig:Models} 
\end{center} 
\end{figure}

Note that the QCD model fitted to the cross section data
\cite{TomasiGustafsson:2005kc}, when extrapolated back to the threshold,
gives a common value for the FFs equal to $ F_{\rm th}\simeq
0.34$. On the other hand, the vector meson dominance (VMD) model of
Ref.~\cite{Bijker:2004yu} gives $ F_{\rm th}\simeq 0.29$. The comparison
between the data and these models is shown in Fig.~\ref{Fig:Models}. The
QCD extrapolation provides, by definition, the same prediction for the two
FFs, as it depends on the number of the quarks involved in the process.
The VDM model of Ref. ~\cite{Bijker:2004yu} predicts a steeper behavior
for $|G_M|$ and a number of resonances occurring in the unphysical
region, i.e., the portion of the TL region lying below the production
threshold. Such a region is accessible through the reaction $\pp\to\epem
\pi^0$~\cite{Adamuscin:2007iv} and can be investigated in next future at
the PANDA@FAIR facility~\cite{Ritman:2005df}.

\section{ Discussion and Conclusions}
We have considered the recent data on TL proton FFs from Ref.
\cite{Ablikim:2019njl,Ablikim:2019eau,CMD-3:2018kql}.
These data confirm the regular oscillations found in Ref.
\cite{Bianconi:2015owa,Bianconi:2016bss}. 
We present a general fit of these data, that includes and updates the previous analysis. 
A more precise determination of the oscillation parameters has been
done. In particular the oscillation period is a relevant parameter since
it has been related to sub-hadron scale processes
\cite{Bianconi:2015owa,Bianconi:2016bss}.  
A similar behavior would be shown by the future $\epem\to n\bar n$ data~\cite{SAhmed:2019}, in this
case the oscillation parameters should bring information on the dynamics
underlying the formation from the vacuum of quark-diquark states, the
quark having different flavor. 

Our analysis does confirm a faster average decrease of the electric FF 
compared to the magnetic one, following a similar behavior as in
the  SL region.  It is in agreement with the predictions of
Ref.~\cite{Kuraev:2011vq}, where such a decreasing behavior was
attributed to the existence of an electrically neutral inner region in
the proton. It is also compatible with the VMD model of
Ref.~\cite{Bijker:2004yu}, that slightly overestimates the magnetic FF.
This appears also  from the fact that the QCD behavior, that does not
differentiate the two FFs, overestimates $G_E$, reproducing better
$G_M$.

The new proton data, together with the future neutron data, will require
a revision of the  phenomenological models based on fitting procedures,
as the parameters were determined in TL region, from the effective FF
only. This will be the object of a future work.

\section{Appendix: expressions of the fit functions}
The change of variables $s=s(p)$, as well as $p=p(s)$, follows from the
relations 
\be 
s=2m_p\left(m_p+\sqrt{p^2+m_p^2}\right)\,,\hspace{10mm}
p=\sqrt{s\left(\frac{s}{4m_p^2}-1\right)}\,. 
\label{eq:eqsp} 
\ee
Therefore Eqs. (\ref{eq:f3p},\ref{eq:fosc}) can be rewritten as:
\ba
F_{\rm 3p}(s)&=& \frac{F_{0}} {\left(
1+\frac{s}{m_a^2}\right)\left(1-\frac{s}{m_0^2} \right)^2}=\nn\\
& =&
\frac{F_{0}} {\left(
1+\frac{2m_p\left(m_p+\sqrt{p^2+m_p^2}\right)}{m_a^2}\right)\left(1- 
\frac{2m_p\left(m_p+\sqrt{p^2+m_p^2}\right)}{m_0^2}\right)^2}
\,,\\
F_{\rm osc}(p(s))&=& 
Ae^{-B\sqrt{s\left(\frac{s}{4m_p^2}-1\right)}}\cos\left[C\sqrt{s\left (
\frac{s}{4m_p^2}-1\right)}+D\right] =Ae^{-Bp}\cos(Cp+D)\,.
\end{eqnarray}

Equation (\ref{eq:FR}) can be expressed as a function of $s$ as
\be 
F_R(\omega(s))=
\displaystyle\frac{1}{1+(\sqrt{s}-
2m_p)^2/r_0} \left
[1+r_1e^{-r_2\left(\sqrt{s}-2m_p\right)}\sin\left(r_3\left(\sqrt{s}-
2m_p\right)\right)\right ]\,, 
\ee 
taking into account that  $\omega=\sqrt{s}-2m_p$. 

\end{document}